\documentclass[showpacs, preprintnumbers, amsmath, amssymb, pre, floatfix, twocolumn]{revtex4-2}
\usepackage{graphicx}
\usepackage{dcolumn}% Align table columns on decimal point
\usepackage{bm}% bold math
\usepackage{amsmath,amsthm,amssymb,ascmac}
\usepackage{color}

\newcommand{\be}{\beta}

\newcommand{\bea}{\begin{eqnarray}}
\newcommand{\eea}{\end{eqnarray}}
\newcommand{\aeq}{&=&}
\newcommand{\aeqe}{&=:&}
\newcommand{\aeqd}{&:= &}
\newcommand{\aeqw}{& \stackrel{\mathrm{w}}{=} &}
\newcommand{\aeqle}{& \le &}
\newcommand{\aeqge}{& \ge &}
\newcommand{\bra}{\langle}

\newcommand{\ket}{\rangle}

\newcommand{\mO}{\mathcal{O}}

\newcommand{\mB}{\mathcal{B}}

\newcommand{\tl}{\tilde}

\newcommand{\defe}{:=}

\newcommand{\ga}{\gamma}
\newcommand{\Ga}{\Gamma}

\newcommand{\al}{\alpha}

\newcommand{\sig}{\sigma}
\newcommand{\f}{\frac}
\newcommand{\half}{\frac{1}{2}}
\newcommand{\pr}{\prime}
\newcommand{\dl}{\delta}
\newcommand{\Dl}{\Delta}

\newcommand{\om}{\omega}

\newcommand{\la}{\label}

\newcommand{\abs}[1]{\vert #1 \vert }
\newcommand{\no}{\nonumber}

\newcommand{\re}[1]{(\ref{#1})}
\newcommand{\res}[1]{\S \ref{#1}}

\newcommand{\hs}{\hspace}

\newcommand{\Bv}[1]{\Big \vert_{#1}}

\newcommand{\RM}[1]{{\rm{#1}}}
\newcommand{\p}{\partial}

\newcommand{\co}[1]{``{#1}''}

\newcommand{\rd}[1]{\textcolor{black}{#1}}
\newcommand{\bl}[1]{\textcolor{black}{#1}}
\begin{document}

\title{Asymptotic expansion of the solution of the master equation and its application to the speed limit}

\author{Satoshi Nakajima}
\email{nakajima@eng.mie-u.ac.jp}
\author{Yasuhiro Utsumi}

\date{\today}

\affiliation{
Department of Physics Engineering, Faculty of Engineering, Mie University, 
Tsu, Mie 514-8507, Japan}
\date{\today}

\begin{abstract}
We investigate an asymptotic expansion of the solution of the master equation under the modulation of control parameters. 
In this case, the non-decaying part of the solution becomes the dynamical steady state expressed as an infinite series using the pseudo-inverse of the Liouvillian, whose convergence is not granted in general. 
We demonstrate that for the relaxation time approximation \bl{model}, the Borel summation of the infinite series is compatible with the exact solution. 
By exploiting the series expansion, we obtain the analytic expression of the heat and the activity. 
In the two-level system coupled to a single bath, under the linear modulation of the energy as a function of time, 
we demonstrate that the infinite series expression is the asymptotic expansion of the exact solution. 
The equality of a trade-off relation between the speed of the state transformation and the entropy production (Shiraishi, Funo, and Saito, Phys. Rev. Lett. ${\bf 121}$, 070601 (2018)) holds in the lowest order of the frequency of the energy modulation in the two-level system. 
To obtain this result, the heat emission and absorption at edges (the initial and end times) or the differences of the Shannon entropy between
the instantaneous steady state and the dynamical steady state at edges are essential: If we ignore these effects, the trade-off relation can be violated.
\end{abstract}

\maketitle 

\section{Introduction} 

Time-dependent open systems have been studied actively in recent years. 
These studies relate to quantum pump \cite{RT09, Flindt10, Splettstoesser12, Yuge12, Nakajima15, Splettstoesser17, Flindt19, Takahashi2020}, 
excess entropy production \cite{Saitou, Sagawa, Komatsu15, Nakajima17, Nakajima17D}, 
efficiency and power of heat engine \cite{Shiraishi16, Brandner, Tajima, Kamimura,Hino2021}, 
shortcuts to adiabaticity~\cite{RMP19, Takahashi17, Funo2020},
and speed limits \cite{Ito18, Shiraishi18, Funo19}. 
In the studies of quantum pumping and excess entropy production for systems governed by the master equation, the time dependence is described using the pseudo-inverse of the Liouvillian \cite{RT09, Flindt10, Splettstoesser12, Nakajima15, Averin17, Splettstoesser17, Nakajima17, Nakajima17D, Hino2021, Flindt19}. 
However, the solution using the pseudo-inverse of the Liouvillian is an asymptotic expansion and does not converge in general: The solution is obtained by iterative applications of the pseudo-inverse and the time derivative, which can make the series expansion divergent similarly to the case for the adiabatic iteration~\cite{Berry1983}. 
\bl{In the two-level system (like spinless one-level quantum dot) coupled to a single bath}, 
the solution expressed by the pseudo-inverse is an asymptotic expansion of the exact solution and provides a good approximation in the first few terms. 

One of the trade-off relations between the speed of state transformation and the entropy production for the classical stochastic process is given in Ref. \cite{Shiraishi18}: 
\rd{An inequality}
\bea
r \defe \f{2\sig A}{L^2} \tau \ge 1 \label{SFS}
\eea
\rd{holds if the Liouvillian satisfies the local detailed balance condition}.
Here, $L$ is the total variation distance \rd{between the states at the initial time $t=0$ and the final time $t=\tau$}, $A$ is the average activity and $\sigma$ is the total entropy production 
(for the definitions, see \res{s_trade_off}. \rd{We derive \re{SFS} in Appendix \ref{A_SFS}}). 
The authors of Ref.~\cite{Shiraishi18} considered a two-level system and provided a specific protocol, which realizes $r=(5/2)\ln(3/2)=1.01366...$ in the limit of slow driving. 
In the present paper, based on the asymptotic expansion of \bl{the two-level system}, 
we will demonstrate protocols that achieve the equality of  \re{SFS}. 

The structure of the paper is as follows. 
First, we give an infinite series expression of the dynamical steady state which is the solution of the master equation under the modulation of control parameters (\res{s_general}). 
Next, in \res{s_exact_N}, we introduce the relaxation time approximation \bl{model (which is analytically tractable)} 
and show that the Borel summation of the infinite series expression becomes the exact solution. 
In \res{s_exact_2}, for \bl{the two-level system}, we consider that linear modulation of the energy as a function of time
and show that the infinite series expression is the asymptotic expansion of the exact solution. 
In \res{s_trade_off}, we show that the equality of \bl{\re{SFS}} holds in the lowest order of the frequency of the energy modulation in the two-level system. 
In \res{s_summary}, we summarize this paper. 
In Appendix \ref{A_SFS}, we derive the trade-off relation of Shiraishi-Funo-Saito. 
In Appendix \ref{A_0}, we derive the exact solution of the master equation.
In Appendix \ref{A_A}, we explain a prescription getting the exact solution from the asymptotic expansion in the two-level system. 
In Appendix \ref{A_HD}, we discuss higher derivative and oscillation.
In Appendix \ref{A_S}, we give an instance of the convergent infinite series expression of the dynamical steady state.
In Appendix \ref{s_heat}, we calculate the heat current and heat using the asymptotic expansion of the dynamical steady state for the relaxation time approximation \bl{model}. 

\section{General theory} \la{s_general}

In this section, we give an infinite series expression of the dynamical steady state using the pseudo-inverse of the Liouvillian. 

We consider a master equation
\bea
\f{d}{dt}p_i(t) \aeq \sum_{j=0}^{N-1} K_{ij}(\al_t)p_j(t). \la{Master_eq}
\eea
Here, $p_i$ is the probability of state $i$($=0,1,\cdots, N-1$) at time $t$ and $\al_t$ is the value of the set of the control parameters at time $t$. 
We assume that the master equation has a unique instantaneous steady state $p_i^\RM{ss}(\al)$ which satisfies
$\sum_{j} K_{ij}(\al)p_j^\RM{ss}(\al)=0 $ and $\sum_i p_i^\RM{ss}(\al)=1$. 
The pseudo-inverse of the Liouvillian $K_{ij}(\al)$ is defined by
\bea
\sum_k R_{ik}(\al)K_{kj}(\al)=\dl_{ij}- p_i^\RM{ss}(\al). \la{def_R}
\eea
Applying the pseudo-inverse $R(\al_t)$ to the master equation \re{Master_eq}, we obtain
\bea
\Big[1-R(\al_t)\f{d}{dt} \Big]\dl p(t) \aeq R(\al_t)\f{dp^\RM{ss}(\al_t)}{dt}, \la{Master_eq_2}
\eea
where $\dl p(t) \defe p(t)-p^\RM{ss}(\al_t)$. 
Here, $R$ denotes the matrix $(R_{ij})$ and $p$ is the vector $(p_0, p_1, \cdots, p_{N-1})^t$. 
The formal solution of \re{Master_eq_2} is given by an infinite series \cite{RT09, Nakajima15}
\bea
\dl \rd{p^\RM{dss}}(t) \aeqd \sum_{n=1}^\infty \Big[ R(\al_t)\f{d }{dt} \Big]^n p^\RM{ss}(\al_t)=:\sum_{n=1}^\infty p^{(n)}(t). \la{def_p^n}
\eea
The general solution of \re{Master_eq_2} is 
\bea
\dl p(t) \aeq \dl \rd{p^\RM{dss}}(t)+\tl \dl p(t).
\eea
Here, $\tl \dl p(t)$ is the solution of
\bea
\Big[1-R(\al_t)\f{d}{dt} \Big]\tl \dl p(t) \aeq 0
\eea
under $\tl \dl p(0)= \dl p(0)-\dl \rd{p^\RM{dss}}(0) $.
$\tl \dl p(t)$ also  satisfies $\f{d}{dt}\tl \dl p(t)=K(\al_t)\tl \dl p(t)$ and damps exponentially as a function of time \cite{Nakajima17D}. 
Then, the general solution of the master equation is given by \cite{Nakajima15, Nakajima17D}
\bea
p(t)=\rd{p^\RM{dss}}(t)+\tl \dl p(t),
\eea
where 
\bea
\rd{p^\RM{dss}}(t) \aeqd p^\RM{ss}(\al_t)+\dl \rd{p^\RM{dss}}(t)\no\\
 \aeq p^\RM{ss}(\al_t)+\sum_{n=1}^\infty p^{(n)}(t) . \la{def_dss}
\eea
We call $\rd{p^\RM{dss}}(t) $  {\it dynamical steady state}. 
The general solution consists of the exponentially damping term $\tl \dl p(t)$ and the dynamical steady state, 
which is composed of the instantaneous steady state and the correction $\dl \rd{p^\RM{dss}}(t)$.

In general, the infinite series $\rd{p^\RM{dss}}(t)$ dose not converge.
In the next section, we compare $\rd{p^\RM{dss}}(t)$ and an exact solution of the master equation. 

\section{Exact solution and Borel  summation} \la{s_exact}

In this section, first, we study the relaxation time approximation \bl{model} for an $N$-level system and show that the Borel summation of the infinite series $\rd{p^\RM{dss}}(t)$ is the exact solution of the master equation. 
Next, we study the two-level system and demonstrate that $\rd{p^\RM{dss}}(t)$ is identical with the asymptotic expansion of the exact solution. 

\subsection{$N$-level system} \la{s_exact_N}

We introduce a Liouvillian
\bea
K_{ij}(\al) \aeq \ga(p_i^\RM{ss}(\al)-\dl_{ij}) \la{K_N}
\eea
with $\ga>0$ \rd{as an analytically tractable model}. 
This Liouvillian satisfies the detailed balance condition $K_{ij}(\al) p_j^\RM{ss}(\al)=K_{ji}(\al) p_i^\RM{ss}(\al)$ \cite{Note1}. 
\rd{This model is not just a toy model becomes it induced the following Liouvillian}:
\bea
K(\al) \aeq \ga \begin{pmatrix} 
-f &&1-f \\
f &&-(1-f) 
\end{pmatrix} ,  \la{K_2}
\eea
\rd{which is the same form with a spinless one-level quantum dot coupled to leads}.
Here, $f=\sum_{b}  \f{\ga_b}{\ga}\f{1}{e^{\be_b \Dl}+1}$ and $\sum_{b}\ga_b=\ga$. 
$\ga_b$ is the coupling strength and $\be_b$ is the inverse temperature of bath $b$.
\rd{$\Dl$ is the Energy level difference between the two levels}. 
The master equation of \re{K_N} is given by
\bea
\f{d}{dt}p_i(t) \aeq -\ga(p_i(t)-p_i^\RM{ss}(\al_t)).
\eea
This is the relaxation time approximation of which relaxation time is $1/\ga$. 
A pseudo-inverse of \re{K_N} is given by
\bea
R_{ij}(\al) \aeq \f{1}{\ga}(p_i^\RM{ss}(\al)-\dl_{ij}) . \la{R_R}
\eea
For simplify,  we assume $\ga$ is a constant in the following of this paper. 
Substituting \re{def_p^n} and \re{R_R} \rd{into} \re{def_dss}, we obtain
\bea
\tl F_i(t) \defe \rd{p^\RM{dss}_i}(t)= \sum_{n=0}^\infty \rd{\f{(-1)^n}{\ga^n} } \f{d^n}{dt^n} p_i^\RM{ss} (\al_t).
\eea
In \res{s_exact_2}, we show that $\tl F_i(t)$ diverges in the two-level system.

We calculate the Borel summation of  $\tl F_i(t)$. 
For a series $\tl S(t) \defe \sum_{n=0}^\infty a_n(t)$, if  (i) the Borel function $\mB[\tl S(t)](s) \defe \sum_{n=0}^\infty \f{a_n(t)}{n!}s^n$
converges to an analytic function for $0 \le s < \rho$ ($\rho>0$) and can be analytically continued along the positive real axis and
(ii) $S(t) \defe \int_0^\infty ds \ e^{-s}\mB[\tl S(t)](s) $ is well defined, $S(t)$ is called as the Borel summation of $\tl S(t)$ \cite{Hardy}. 
The Borel function of $\tl F_i(t)$ is given by
\bea
\mB[\tl F_i(t)](s) \aeq  \sum_{n=0}^\infty \f{1}{n!}s^n \rd{\f{(-1)^n}{\ga^n} } \f{d^n}{dt^n} p_i^\RM{ss}(t) ,
\eea
where $p_i^\RM{ss}(t)$ denotes $p_i^\RM{ss}(\al_t)$. 
This series converges to an analytic function $p_i^\RM{ss}(t-\f{s}{\ga})$ for small $s$. 
Then, the Borel summation of $\tl F_i(t)$ is given by 
\bea
\int_0^\infty ds \ e^{-s}\mB[\tl F_i(t)](s) \aeq \int_0^\infty ds \ e^{-s} p_i^\RM{ss}\Big(t-\f{s}{\ga}\Big) \no\\
\aeqe F_i(t). \la{def_F_i}
\eea
 
Next, we show that the exact solution of the master equation is identical with the Borel summation of $\tl F_i(t)$. 
Using the general theory of first-order linear ordinary differential equation, the exact solution under the initial condition $p_i(t=0)=p_i(0)$ is given by
\bea
p_i(t) \aeq e^{-\ga t} \Big[p_i(0)+\ga \int_0^t du \ p_i^\RM{ss}(u)e^{\ga u} \Big]. \la{exact_pre}
\eea
This equation can be rewritten as (Appendix \ref{A_0})
\bea
p_i(t) \aeq e^{-\ga t} [p_i(0)-F_i(0)]+F_i(t), \la{exact_p}
\eea
where $F_i(t)$ was introduced in \re{def_F_i}. 
The first term of \re{exact_p} corresponds to $\tl \dl p(t)$ because this term damps exponentially. 
The second term of \re{exact_p} corresponds to the dynamical steady state.
We note that the exact solution of the master equation is identical with the Borel summation of {$\rd{p_i^\RM{dss}}(t)$ even if $\ga$ is time dependent.

In the following of this paper, we assume that the instantaneous steady state $p^\RM{ss}(\al)$ is the Gibbs distribution $p^\RM{ss}(\al)=e^{-\be E_i}/Z$ where 
$E_i$ is the energy of the state $i$, $\be$ is the inverse temperature and $Z \defe \sum_i e^{-\be E_i} $ is the partition function. 
We focus on the energy modulation 
\bea
E_i = \al(t) E_i^{(0)} \la{E(t)}
\eea
with fixed $E_i^{(0)}$ and $\beta$ and study the linear modulation $\al(t) = h_\RM{i} + \om t$ in particular. 
In this case, the control parameter is $\al(t)$.

\subsection{Demonstration in two-level system}  \la{s_exact_2}

In this subsection, we consider the two-level system \re{K_2} and show that $\tl F_i(t)$ $(i=0,1)$ diverges and is identical with the asymptotic expansion of the exact solution $F_i(t)$. 

We consider \re{K_2} with $\Dl(t)=(h_\RM{i}+\om t)\Dl$ and fixed $\ga$ and $\beta$ in the following of this section. 
In this case, $\tl F(t) \defe \tl F_1(t)$ is given by
\bea
\tl F(t) =\sum_{n=0}^\infty \rd{\f{(-1)^n}{\ga^n}} \f{d^n}{dt^n}  f(W+Vt)
\eea
with $f(x) \defe 1/(e^x+1)$. 
Here, $W\defe \be \Dl  h_\RM{i}$, $V \defe \be \Dl \om>0$. 
Figure \ref{Fig_1} shows $F(t)(\defe F_1(t))$ and $\tl F_{(N)}(t) \defe \sum_{n=0}^N \rd{\f{(-1)^n}{\ga^n}} \f{d^n}{dt^n} f(W+Vt)$ for 
$\bl{N=0}$, 5, 24, 25, and 26 at $V/\ga=0.3$. 
$\tl F(t)$ dose not converge. 
For $N=5$, the asymptotic expansion $\tilde{F}_{(5)}$ well agrees with the exact solution $F(t)$. 
However, for large $N=24$, $25$ and $26$, the deviations increase and the asymptotic expansions oscillate. 
\bl{Although probability should be between 0 and 1, $\tl F_{(26)}(t)$ is not}.

\begin{figure}
\includegraphics[width=1 \columnwidth]{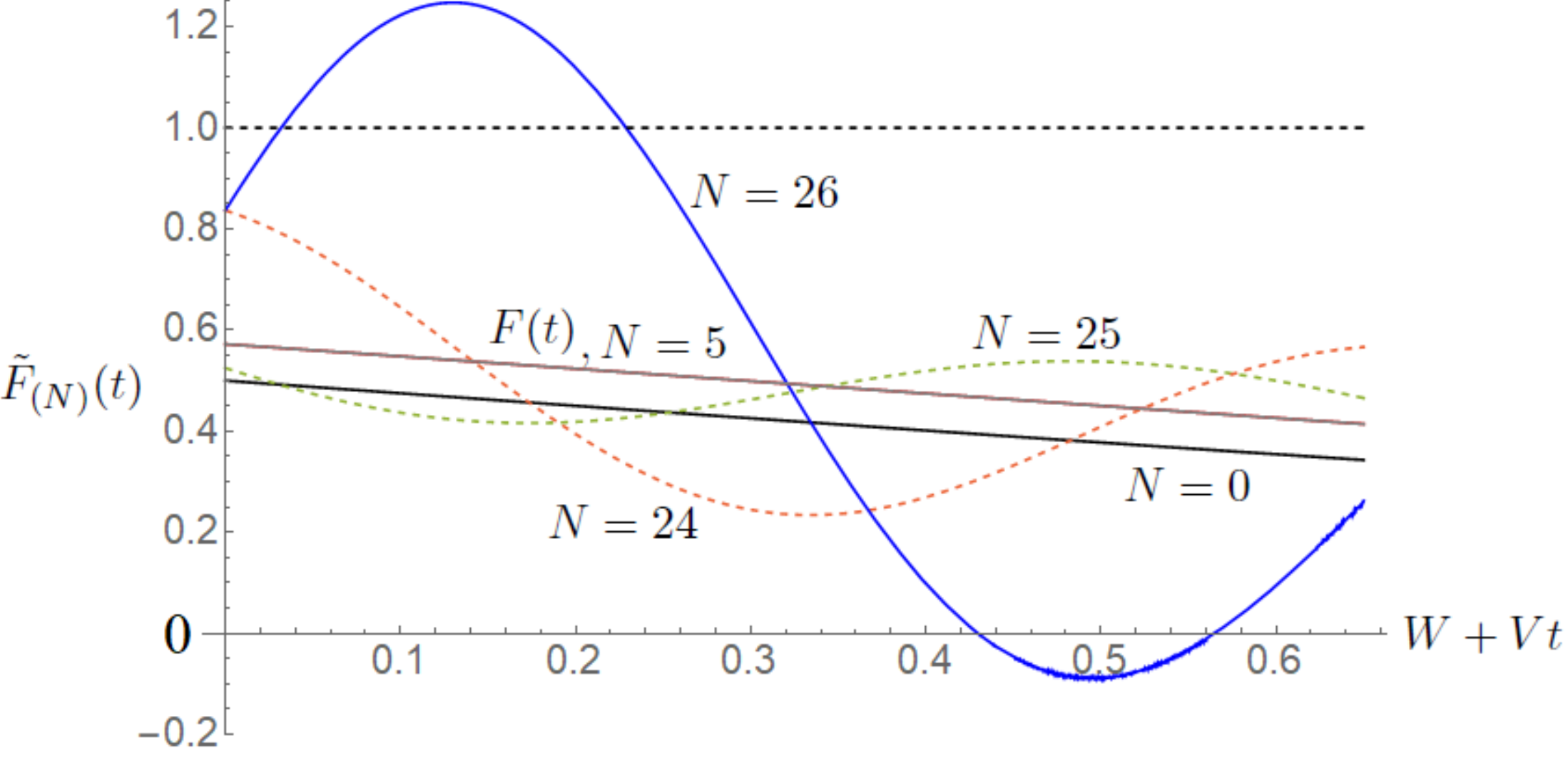}
\caption{\label{Fig_1}$F(t)(\defe F_1(t))$ and $\tl F_{(N)}(t)$ for $\bl{N=0}$, 5, 24, 25, and 26 at $V/\ga=0.3$. 
$F(t)$ and $\tl F_{(5)}(t)$ are overlapping. 
The asymptotic expansion above 24th order deviates significantly from the exact solution. }
\end{figure}

The dynamical steady state is given by
\bea
F(t) \aeq  \int_0^\infty ds \ e^{-s}f\Big(W+V\Big(t-\f{s}{\ga} \Big) \Big) \no\\
\aeq {}_2F_1(1,\bl{T};\bl{T}+1;-e^{W+Vt}) \no\\
\aeq f(W+Vt) {}_2F_1(1,1;\bl{T}+1;1- f(W+Vt)) \la{def_F}
\eea
with $\bl{T} \defe \ga/V$. 
${}_2 F_1(a,b;c;z)$ is the hypergeometric function. 
In the third equality of \re{def_F}, we used the Kummer's relation ${}_2F_1(a,b;c;z) = \f{1}{(1-z)^a}{}_2F_1(a,c-b;c;\f{z}{z-1})$.

We show that $\tl F(t)$ is an asymptotic expansion of $F(t)$. 
A divergent series $\sum_{k=0}^\infty a_k/\bl{T}^k$ is the asymptotic expansion of a function 
$G(\bl{T})$ if $R_n(\bl{T}) \defe \bl{T}^n[G(\bl{T})- \sum_{k=0}^n a_k/\bl{T}^k]$ satisfies $\abs{R_n(\bl{T})} \to 0 \  (\bl{T} \to \infty)$. 
In the third line of \re{def_F}, $ {}_2F_1(1,1;\bl{T}+1;z) $ can be expanded in $z$ as
\bea
\hs{-3mm}&& \hs{-8mm} {}_2F_1(1,1;\bl{T}+1;z) \no\\
\hs{-3mm}\aeq 1+\sum_{k=1}^\infty \f{k!}{(\bl{T}+1)(\bl{T}+2)\cdots (\bl{T}+k)} z^k \no\\
\hs{-3mm} \aeqe 1+\sum_{k=1}^n \f{k!}{(\bl{T}+1)(\bl{T}+2)\cdots (\bl{T}+k)} z^k + \dl_n(\bl{T},z). \la{eq22}
\eea
The second term of the third line of \re{eq22} can be expanded in $1/\bl{T}$ as 
\bea
&& \hs{-5mm} \sum_{k=1}^n \f{k!}{(\bl{T}+1)(\bl{T}+2)\cdots (\bl{T}+k)} z^k \no\\
\aeq \sum_{k=1}^n \f{a_k(z)}{\bl{T}^k}+r_n(\bl{T}, z)
\eea
with $r_n(\bl{T}, z) = \mO(1/\bl{T}^{n+1})$. 
Here, $a_1(z)=z$, $a_2(z)=-z+2z^2$, $a_3(z)=z-6z^2+6z^3$, $a_4(z)=-z+14z^2-36z^3+24z^4$ and $a_5(z)=z-30z^2+150z^3-240z^4+120z^5$. 
Then, $ {}_2F_1(1,1;\bl{T}+1;z) $ is given by 
\bea
 {}_2F_1(1,1;\bl{T}+1;z) \aeq  1+ \sum_{k=1}^n \f{a_k(z)}{\bl{T}^k}+\f{R_n(\bl{T}, z)}{\bl{T}^n}
\eea
with $R_n(\bl{T}, z) \defe \bl{T}^n[r_n(\bl{T}, z)+\dl_n(\bl{T},z)]$. 
The first and second terms of the right hand side provide $\tl F_{(n)}(t)$. 
If $\bl{T}$ is large enough, ${}_2F_1(1,1;\bl{T}+1;z)$ ($0\le z \le 1$) is well approximated by first few terms.
The truncation error $\dl_n(\bl{T}, z) $ satisfies 
\bea
\abs{\dl_n(\bl{T}, z)} &\le& \dl_n(\bl{T}, 1) \no\\
\aeq \f{(n+1)!}{(\bl{T}-1)(\bl{T}+1)(\bl{T}+2)\cdots (\bl{T}+n)}.
\eea
Here, we used the Gauss's summation theorem ${}_2F_1(a,b;c;1)= \f{\Ga(c)\Ga(c-1)}{\Ga(c-a-b)\Ga(c-b)}$ $(\RM{Re}(c)>0$, $\RM{Re}(c-a-b)>0) $
where $\Ga(z)$ is the gamma function. 
The above inequality leads \rd{to} $\abs{R_n(\bl{T}, z)} \to 0\ (\bl{T} \to \infty)$. 
Then, $\tl F(t)$ is the asymptotic expansion of $F(t)$. 
We obtain $\tl F(t)$ from $F(t)$ if we expand ${}_2F_1(1,1;\bl{T}+1;z)$ regarding $\bl{T}$ is \rd{larger} than any natural number $n$.
In addition, we explain a prescription getting $F(t)$ from $\tl F(t)$ in Appendix \ref{A_A}. 

The oscillation of $\tl F_i(t)$ is not limited to two level-system. 
In Appendix \ref{A_HD}, we discuss higher derivative and oscillation for a wide class of functions. 

We note that any infinite series expansion \re{def_dss}  is not necessarily divergent. 
For instance, the protocol discussed in Ref.~\cite{Shiraishi18} leads \bl{to} a convergent series expansion. 
In Appendix \ref{A_S}, we give an instance of the convergent infinite series expression of the dynamical steady state.

\section{Trade-off relation} \la{s_trade_off}

In this section, we show that the equality of a trade-off relation of Shiraishi-Funo-Saito \cite{Shiraishi18} holds
in the lowest order of the frequency of the energy modulation in the two-level system. 

If the Liouvillian satisfies the local detailed balance condition, \bl{the trade-off relation \re{SFS}} holds \cite{Shiraishi18}. 
There, $L \defe \sum_i \abs{p_i(\tau)-p_i(0)}$ is the $l_1$ norm \cite{Cover_Thomas}. 
The average activity
\bea
A \defe \f{1}{\tau}\int_0^\tau dt \ A(t) \la{activity} 
\eea
 is calculated from the activity 
\bea
A(t) \defe \sum_{i\ne j} K_{ij}(t)p_j(t).
\eea
The total entropy production $\sig$ is given by
\bea
\sig \aeqd \be Q +S_\RM{Sh}(p(\tau))-S_\RM{Sh}(p(0)) , \la{sig}
\eea
where $Q$ is the heat flowing into the bath from the system and 
$S_\RM{Sh}(p) \defe -\sum_i p_i \ln p_i$ is the Shannon entropy of the system.

We consider three protocols (Fig. \ref{Fig_3}).
In the following, we suppose that $\al(t)$ is constant for $t \le -\dl/2$ and $t\ge \tau +\dl/2$, and $\al(t)=h_\RM{i}+\om t$ while $-\dl/2 < t < \tau + \dl/2$, 
where $\dl$ is a small value:
\bea
\al(t) \aeq \left \{ \begin{array}{ll}
h_\RM{i}-\om \f{\dl}{2} \hs{14.3 mm} \big(t\le -\f{\dl}{2}\big)  \\
h_\RM{i}+\om t  \hs{15.7 mm} \big(-\f{\dl}{2} \le t \le \tau +\f{\dl}{2} \big) \\ 
h_\RM{i}+\om \big(\tau+\f{\dl}{2} \big) \hs{5 mm} \big(\tau +\f{\dl}{2} \le t \big)  \\
\end{array} \right. . \la{al_ABC}
\eea
In the first protocol (protocol A), we consider $-\dl \le t \le \tau+\dl $. 
At time  $t=-\dl$ and $t=\tau+\dl$, the state of the system is the instantaneous steady state $p^\RM{ss}$.
In the second protocol (protocol B), we consider $0 \le t \le \tau$ and suppose that the state of the system at $t=0$ and $t=\tau$ is the dynamical steady state $\rd{p^\RM{dss}}$.
In the third protocol (protocol C), we consider $-\dl \le t \le \tau$ and suppose that the state of the system at $t=-\dl$ is the instantaneous steady state. 
\rd{Protocols B and C can be regarded as observations of  the process of protocol A at different time intervals.}
The ratio $r$ defined by \re{SFS} can be expanded as
\bea
r = r_0 + r_1 \om + r_2 \om^2 +\cdots .
\eea
In the following, we show that $r_0=1$ for protocol A, B, and C in the two-level system \re{K_2}.
To calculate $r_0$, we estimate $\sig$ up to first order of $\om$ (we denote this by $\sig^{(1)}$) and the average activity $A$ and $L$ up to zeroth order. 
In the zeroth order of $\om$, $A$ is given by $A^\RM{ss} \defe \f{1}{\tau}\int_0^\tau dt \ A^\RM{ss}(\al_t)$ with $A^\RM{ss}(\al_t) \defe \sum_{i\ne j} K_{ij}(\al_t)p_j^\RM{ss}(\al_t)$, 
and $L$ is given by $L^\RM{ss} \defe \sum_i \abs{p^\RM{ss}_i(\al_\tau)-p^\RM{ss}_i(\al_0)}$. 

\begin{figure}
\includegraphics[width=1 \columnwidth]{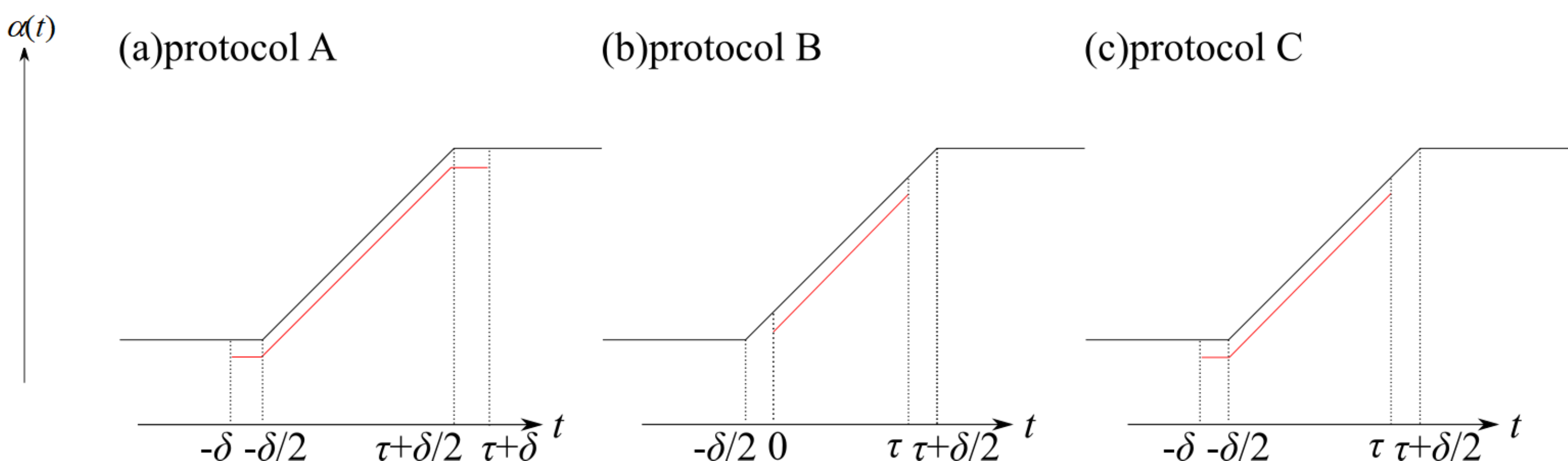}
\caption{\label{Fig_3} Protocol A, B, and C.}
\end{figure}

We estimate $A^\RM{ss}$ in \rd{the} relaxation time approximation \bl{model} \re{K_N}.
In this case, we obtain
\bea
A^\RM{ss}(\al_t) = \ga \Big[ 1 - \sum_i (p_i^\RM{ss})^2 \Big].
\eea
%The right hand side is related to the purity \cite{Nielsen_Chuang}. 
In the two-level system \re{K_2} with $\Dl(t)=\al(t)\Dl$, $A^\RM{ss}$ becomes
\bea
A^\RM{ss} \aeq \f{2\ga}{\Dl(\be_\RM{f}-\be_\RM{i})} \Big[f(\be_\RM{i} \Dl)-f(\be_\RM{f} \Dl)  \Big] .\la{A_ss}
\eea
Here, $\be_\RM{i} \defe h_\RM{i} \be$, $\be_\RM{f} \defe h_\RM{f} \be$, and $h_\RM{f} \defe h_\RM{i}+\om \tau$.

We evaluate the heat current 
\bea
q(t) \defe -\sum_i E_i \f{d}{dt} p_i \la{def_q(t)}
\eea
using the asymptotic expansion \re{def_dss}.
We denote the heat current contributed from $ p^\RM{ss}(\al_t)$ and $p^{(n)}(t)$ by $q_0(t)$ and  $q_{n}(t)$ respectively.
We obtain 
\bea
q_0(t) \aeq -\sum_i E_i \f{dp_i^\RM{ss}}{dt}, \la{q_0_def}\\
q_1(t) \aeq -\sum_{i,k} E_i \Big[R_{ik}\f{d^2p_k^\RM{ss}}{dt^2} -\Big(R\f{dK}{dt}R\Big)_{ik} \f{dp_k^\RM{ss}}{dt} \Big] \la{q_1_general}.
\eea
Substituting \re{E(t)} \bl{into} \re{q_0_def}, we obtain
\bea
q_0(t) \aeq \al^\pr(t) \al(t) \be [\bra (E^{(0)})^2 \ket- \bra E^{(0)} \ket^2],
\eea
where $\bra (E^{(0)})^n \ket \defe \sum_i (E_i^{(0)})^n p_i^\RM{ss}(\al_t)$ is the $n$-th moment of energy.
$q_0(t)$ is proportional to the variance of the energy and does not depend on the Liouvillian.
It is also expressed by using the Fisher information \cite{Ito18, Cover_Thomas, Note2}. 
For \re{al_ABC}, the zeroth order of the heat $Q_0 = \int_0^\tau dt \ q_0(t)$ is given by
\bea
Q_0 \aeq \f{1}{\be}\big[ \be_\RM{i} \bra E^{(0)} \ket(\be_\RM{i}) + \ln Z(\be_\RM{i}) \no\\
&&-\be_\RM{f} \bra E^{(0)} \ket(\be_\RM{f}) - \ln Z(\be_\RM{f}) \big] .
\eea
The Shannon entropy is given by $S_\RM{Sh}(p) = S_\RM{Sh}(\be_t) + \dl S_\RM{Sh}$ with $S_\RM{Sh}(\be_t) \defe \be_t \bra E^{(0)} \ket(\be_t) + \ln Z(\be_t)$ and $\be_t \defe \al(t)\be$.
Therefore, the zeroth order of the entropy production vanishes:
$\be Q_0 + S_\RM{Sh}(\be_\RM{f}) -S_\RM{Sh}(\be_\RM{i}) =0$.
For the relaxation time approximation \bl{model}, $q_1(t)$ is given by 
\bea
q_1(t) \aeq \f{\al(t)}{\ga}\Big( [\al^\pr(t)]^2 \be^2 C_3(\be_t) -\al^{\pr\pr}(t)\be C_2(\be_t)  \Big). \la{q_1_R}
\eea
Here, $C_n(\be_t)$ is $n$-th order cumulant of $E_i^{(0)}$:
\bea
C_n(\be_t)=(-1)^n \f{\p^n \ln Z(\be_t)}{\p \be_t^n}
\eea
and $Z(\be_t) \defe \sum_i e^{-\be_t E_i^{(0)}} $. 
In Appendix \ref{s_heat}, we calculate $q_n(t)$.

We calculate $\sig^{(1)}$ for the relaxation time approximation \bl{model}.
For each protocols, $\sig^{(1)}$ is given by
\bea
\sig^{(1)} \aeq \be \Big(Q_1+\dl Q_1 \Bv{t=\tau} - \dl Q_1\Bv{t=0} \Big)  \ \ (\mbox{protocol A}) ,\no\\
\sig^{(1)} \aeq \be Q_1 + \dl S_\RM{Sh}\Bv{t=\tau} -  \dl S_\RM{Sh}\Bv{t=0} \ \ (\mbox{protocol B}) ,\no\\
\sig^{(1)} \aeq \be \Big(Q_1 - \dl Q_1\Bv{t=0} \Big) +  \dl S_\RM{Sh}\Bv{t=\tau}  \ \ (\mbox{protocol C}). \no\\
\eea
Here, 
\bea
Q_1 \aeqd \int_0^\tau dt \ q_1(t) \no\\
\aeq \f{\om }{\ga} \big[ \be_\RM{i} C_{2}(\be_\RM{i}) +C_{1}(\be_\RM{i})  -\be_\RM{f} C_{2}(\be_\RM{f}) -C_{1}(\be_\RM{f}) \big] .
\eea
$\dl Q_1$ is the heat contributed from the term ($\al^{\pr\pr}(t)$) including the delta function in \re{q_1_R} and given by
\bea
\dl Q_1 \aeq \f{\om }{\ga}\be_t C_2(\be_t) .
\eea
$\dl S_\RM{Sh}$ is given by 
\bea
\dl S_\RM{Sh} \aeq - \sum_{i} p_i^{(1)}\ln p_i^\RM{ss} \no\\
\aeq \f{\om}{\ga} \be \be_t C_2(\be_t) .
\eea
Then,  $\sig^{(1)}$ becomes
\bea
\sig^{(1)} =\f{\om}{\ga} \be [C_1(\be_\RM{i}) - C_1(\be_\RM{f})]  \la{sig_ABC_R}
\eea
for protocol A, B, and C.
In the two-level system, 
\bea
\sig^{(1)} =\f{\om}{\ga} \be \Dl [f(\be_\RM{i} \Dl)-f(\be_\RM{f} \Dl)]  \la{sig_ABC}
\eea
holds. 

From \re{A_ss}, \re{sig_ABC} and
\bea
L^\RM{ss} \aeq 2[f(\be_\RM{i} \Dl)-f(\be_\RM{f} \Dl)],
\eea
we obtain 
\bea
r_0 = \f{2\sig^{(1)} A^\RM{ss}}{(L^\RM{ss})^2}\tau =1 \la{r_0=1}
\eea
for protocol A, B, and C.
To obtain \re{r_0=1}, the heat emission and absorption at edges \bl{($t=-\dl/2$, $\tau+\dl/2$)} 
or the correction of the Shannon entropy $\dl S_\RM{Sh}$ at edges \bl{($t=0$, $\tau$)} are essential.
If we ignore these corrections, $r_0$ can be less than 1.

\begin{figure*}
\includegraphics[width=2 \columnwidth]{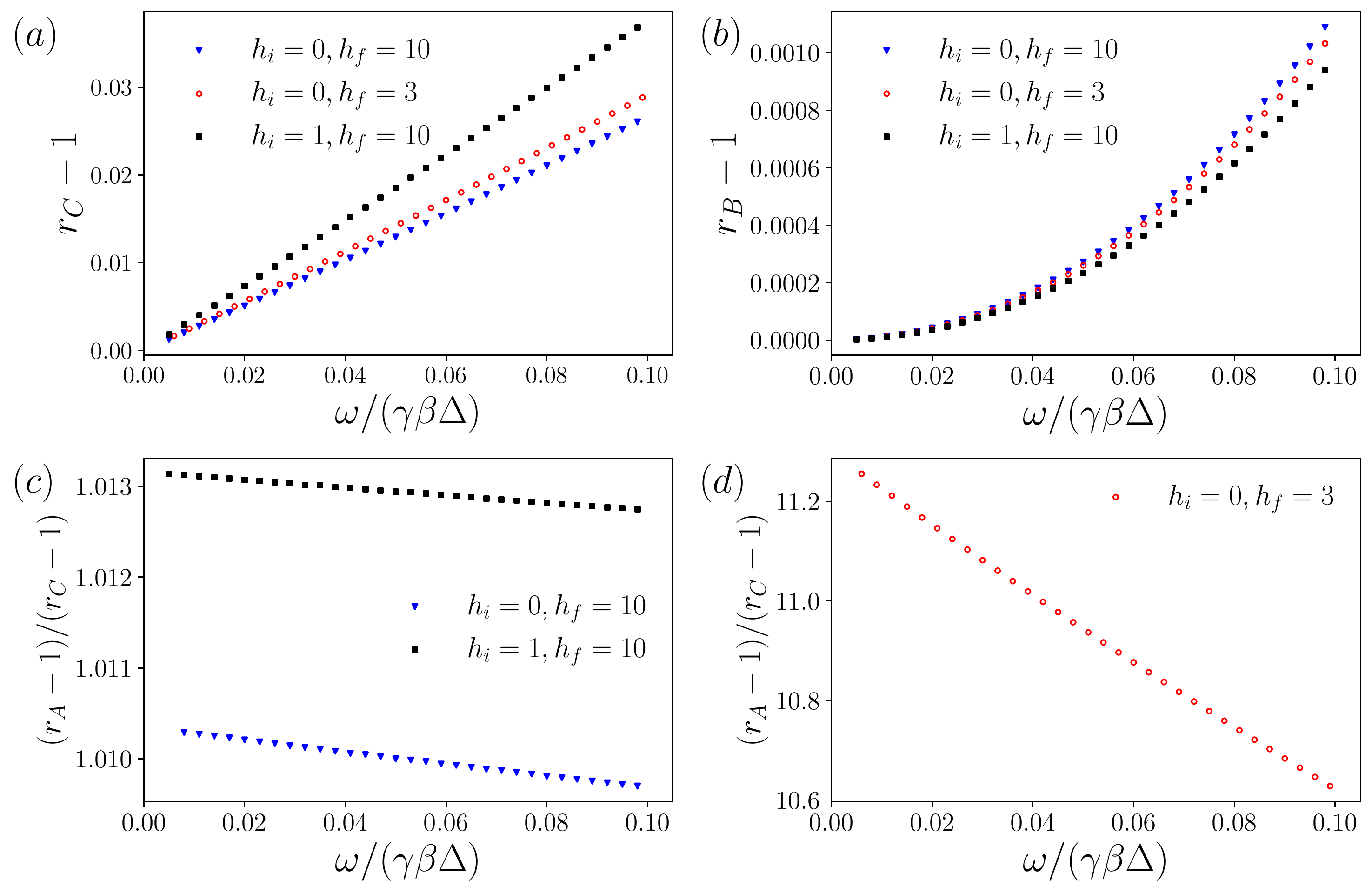}
\caption{\label{Fig_2}$(a)$ $(r-1)$ for protocol C, $(b)$ $(r-1)$ for protocol B, 
(c) $(r_A-1)/(r_C-1)$ and $(d)$ $(r_A-1)/(r_C-1)$ for $(h_\RM{i},h_\RM{f})=(0,10)$ (triangle), $(h_\RM{i}, h_\RM{f})=(1,10)$ (square) and $(h_\RM{i},h_\RM{f})=(0,3)$ (circle).  
Here, $r_A$ and $r_C$ denote $r$ for protocol A and C respectively. 
In these figures, we set $\be \Dl =1$. }
\end{figure*}

Figure \ref{Fig_2}, shows the numerical results of $(r-1)$ for the three protocols calculated from \re{activity} and \re{def_q(t)} by exploiting the exact solution for the two-level system \re{def_F}. 
For all protocols, during the duration $0 \le t \le \tau$, the system is driven as $\al(t)=h_\RM{i}+\om t$. 
The initial states for the protocols A, B and C are the instantaneous steady state $p_1(t=0)=f(W)$, the dynamical steady state $p_1(t=0)=F(0)$ and the instantaneous steady state $p_1(t=0)=f(W)$, respectively. 
In addition, for the protocol A, after we turn off the driving at time $\tau$, we fix $\al(t)=h_\RM{f}=h_\RM{i}+\om \tau$ ($\tau \le t \le \tau+30/\ga$) and wait until the system relaxes to the instantaneous steady state. 
In this duration the heat is emitted from the system.

In Fig. \ref{Fig_2}, we plot $(r_C-1)$, $(r_B-1)$ and $(r_A-1)/(r_C-1)$ for various $h_\RM{i}$ and $h_\RM{f}$: $(h_\RM{i},h_\RM{f})=(0,10)$ (triangle), $(h_\RM{i}, h_\RM{f})=(1,10)$ (square) and $(h_\RM{i},h_\RM{f})=(0,3)$ (circle) at $\be \Dl =1$.
Here, $r_A$ and $r_C$ denote $r$ for protocol A and C respectively. 
Figure \ref{Fig_2} shows that $r_0=1$, $r_1>0$ for protocol A and C and $r_0=1$, $r_1=0$, $r_2>0$ for protocol B. 
Namely, for small $\om$, the correction $(r-1)$ is a linear function for protocol A and C and a quadratic function for protocol B. 

In systems described by the master equation having the Liouvillian \re{K_N} or more general Liouvillians, in general, 
$r_0 > 1$ if $N \ge 3$. 
For instance, in the relaxation time approximation \bl{model},  $r_0$ of protocol A of a $N$-level ($N \ge 2$) system of which energies are $E_0^{(0)}=0$ and $E_i^{(0)}=\Dl$ ($i=1,2, \cdots, N-1)$ is given by 
\bea
r_0 \aeq \f{N}{2(N-1)} \no\\
&&+\f{N-2}{2(N-1)^2}\f{\ln \f{1+(N-1)e^{-\be_\RM{i}\Dl}}{1+(N-1)e^{-\be_\RM{f}\Dl}}}{\f{1}{N-1+e^{\be_\RM{i}\Dl}}-\f{1}{N-1+e^{\be_\RM{f}\Dl}}}  .
\eea

\section{Summary} \la{s_summary}

In summary, for the relaxation time approximation \bl{model}, we calculated the infinite series expansion of the solution of driven master equation using the pseudo-inverse of the Liouvillian. 
We demonstrated that the Borel summation of the series becomes the exact solution. 
For the two-level system coupled to a single bath, we considered that the linear modulation of the energy
and showed that the series expression is the asymptotic expansion of the exact solution. 
Based on the asymptotic expansion for the relaxation time approximation \bl{model}, we demonstrated that the equality of the trade-off relation of Shiraishi-Funo-Saito \cite{Shiraishi18} 
holds in the lowest order of the frequency of the energy modulation in the two-level system. 
To our knowledge, this is the first  instance where the equality sign holds in the trade-off relation of Ref. \cite{Shiraishi18}.
If we ignore the heat emission and absorption at edges (the initial and end times) or 
the differences of the Shannon entropy between the instantaneous steady state and the dynamical steady state at edges, the trade-off relation can be broken. 

\acknowledgments

We acknowledge helpful discussions with Y. Tokura and Y. Teratani. 
This work was supported by JSPSKAKENHI Grants No. 18KK0385, and No. 20H01827. 

\appendix

\section{Derivation of the trade-off relation} \la{A_SFS}

To be self contained, we repeat the derivations of \re{SFS} for single heat bath case (Ref. \cite{Shiraishi18} studied multiple heat baths case).
Suppose the local detailed balance condition
\bea
K_{ij}(t)e^{-\be(t)E_j(t)}=K_{ji}(t)e^{-\be(t)E_i(t)} .\la{LDB}
\eea
Here,  $\be(t)$ is the inverse temperature of the bath. 
The entropy production rate $\dot{\sig}(t)$ is defined by 
\bea
\dot{\sig}(t) \defe \f{d}{dt}S_\RM{Sh}(p(t)) +\be(t)q(t). \la{def_dot_sig}
\eea
The heat current $q(t)$ defined by \re{def_q(t)} can be rewritten as
\bea
q(t) \aeq \sum_{i\ne j}K_{ji}(t)p_i(t) [E_i(t)-E_j(t)]. \la{heat_current_A}
\eea
Here, we used  the master equation 
\bea
\f{dp_i}{dt}=\sum_{j(\ne i)}[K_{ij}(t)p_j(t) - K_{ji}(t)p_i(t)]. \la{ME_A}
\eea
Using \re{heat_current_A}, \re{ME_A} and the local detailed balance condition, \re{def_dot_sig} becomes
\bea
\dot{\sig}(t) \aeq \half  \sum_{i \ne j} (K_{ji}p_i-K_{ij}p_j)\ln \f{K_{ji}p_i}{K_{ij}p_j} \no\\
\aeqge \sum_{i \ne j}\f{(K_{ji}p_i-K_{ij}p_j)^2}{K_{ji}p_i+K_{ij}p_j}.
\eea
Here, we used the inequality $(a-b)\ln \f{a}{b} \ge 2\f{(a-b)^2}{a+b}$, which is valid for non-negative $a$ and $b$.
The distance $L = \sum_i \abs{p_i(\tau)-p_i(0)}$ is evaluated as
\bea
L \aeqle  \int_0^\tau dt \ \sum_i\Big \vert \f{dp_i}{dt}\Big \vert .
\eea
Here, 
\bea
&&\hs{-5mm}\sum_i \Big \vert \f{dp_i}{dt}\Big \vert = \sum_i \Big \vert \sum_{j(\ne i)} (K_{ij}p_j-K_{ji}p_i) \Big \vert \no\\
\aeqle \sum_i  \sum_{j(\ne i)} \Big \vert K_{ij}p_j-K_{ji}p_i \Big \vert \no\\
\aeqle \sum_{i} \sqrt{\Big(\sum_{j(\ne i)} \f{(K_{ij}p_j-K_{ji}p_i)^2}{(K_{ij}p_j+K_{ji}p_i)} \Big) \sum_{j(\ne i)} (K_{ij}p_j+K_{ji}p_i) } \no\\
\aeqle \sqrt{\Big(\sum_{j\ne i} \f{(K_{ij}p_j-K_{ji}p_i)^2}{(K_{ij}p_j+K_{ji}p_i)} \Big)  \sum_{j\ne i} (K_{ij}p_j+K_{ji}p_i) } \no\\
\aeqle \sqrt{2\dot{\sig}(t) A(t) } \la{der}
\eea
holds. Here, we used the Schwarz inequality in the third and fourth line.
Then, we obtain
\bea
L \aeqle  \int_0^\tau dt \ \sqrt{2\dot{\sig}(t) A(t) } \le \sqrt{2\sig \tau A}. \la{der2}
\eea
Here, we used $\sig \defe \int_0^\tau dt \ \dot{\sig}(t)$, \re{activity} and the Schwarz inequality.
The above equation leads \bl{to} \re{SFS}. 
If $\be(t)$ is time-independent, $\sig$ is given by \re{sig}. 
 
 If
 \bea
 &&\hs{-5mm}\abs{K_{ij}(t)p_j(t)-K_{ji}(t)p_j(t)} \no\\
\aeq \sqrt{\f{\sig}{2\tau A}}[K_{ij}(t)p_j(t)+K_{ji}(t)p_i(t)] \la{der3}
 \eea
holds for all $i \ne j$, the inequalities for the third and fourth  line in \re{der} hold exactly and the second inequality in \re{der2} 
is satisfied with a relative error $\mO(\om^2)$ where $\om$ is the modulation frequency. 
For the two-level system in \res{s_trade_off}, the above equation is satisfied up to the first order of $\om$ while $0\le t \le \tau$. 
The equality of the last line in \re{der} holds with a relative error $\mO(\om^2)$.

\section{Derivation of \re{exact_p}} \la{A_0}

By variable transformation $s=\ga(t-u)$, \re{exact_pre} becomes
\bea
p_i(t)
&=& e^{-\ga t} p_i(0)+ \int_0^\infty ds \ p_i^\RM{ss}\Big(t-\f{s}{\ga}\Big)e^{-s} \no\\
&&- \int_{\ga t}^\infty ds \ p_i^\RM{ss}\Big(t-\f{s}{\ga}\Big)e^{-s} .
\eea
Using variable transformation $-\f{s^\pr}{\ga}=t-\f{s}{\ga}$ in the last term, we obtain \re{exact_p}.

\section{A prescription getting $F(t)$ from $\tl F(t)$} \la{A_A}

We can obtain $F(t)$ from $\tl F(t)$ by following prescription. 
Because of the analytic continuation, it is enough to consider $\abs{e^{W+Vt}}<1$ case.
Then, $F(t)$ is given by 
\bea
F(t) \aeq {}_2F_1(1,\bl{T};\bl{T}+1;-e^{W+Vt})\no\\
\aeq \sum_{n=0}^\infty \f{1}{1+n/\bl{T}}(-e^{W+Vt})^n .
\eea
On the other hand, $\tl F(t)$ is expressed as 
\bea
\tl F(t) \aeq \sum_{m=0}^\infty  \Big(-\f{1}{\ga}\Big)^m \f{d^m}{dt^m} \sum_{n=0}^\infty (-e^{W+Vt})^n  \no\\
\aeqw \sum_{n=0}^\infty \sum_{m=0}^\infty  \Big(-\f{1}{\ga}\Big)^m (nV)^m  (-e^{W+Vt})^n \no\\
\aeq  \sum_{n=0}^\infty \Big[\sum_{m=0}^\infty  \Big(-\f{n}{\bl{T}}\Big)^m \Big] (-e^{W+Vt})^n \no\\
\aeqw  \sum_{n=0}^\infty \f{1}{1+n/\bl{T}}(-e^{W+Vt})^n .
\eea
In the second line, we changed the order of the sum, and in the fourth line, we suppose that $\bl{T}>n$ for all $n$. 
Then, $\tl F(t)$ becomes $F(t)$.

\section{Higher derivative}  \la{A_HD}

$f(W+Vt)$ can be written as
\bea
f(W+Vt) \aeq \half  - \f{1}{V}\sum_{k=0}^\infty \Big[ \f{1}{t-[-W/V+i(\pi/V)(2k+1)]} \no\\
&& + \f{1}{t-[-W/V-i(\pi/V)(2k+1)]}\Big].
\eea
The dominant contribution of $\f{d^n}{dt^n}f(W+Vt)$ (for large $n$) comes from two poles ($k=0$) most nearly above and below $t$. 
In general, $n$-th derivative of $g(t)=\sum_k A_k/(t-a_k)$ ($a_k$ are complex numbers) is given by 
\bea
g^{(n)}(t)=(-1)^n n! \sum_k \f{A_k }{(t-a_k)^{n+1}}.
\eea
The factorial increase with $n$ of the coefficients underlies the divergence commonly encountered in asymptotic series, and, together with the increasingly fast oscillations, reflects the instability of differentiation \cite{Berry}. 
In the studies of the full-counting statistics, the oscillations coming from the higher derivative by the counting field have been observed \cite{Exp, Flindt10, Utsumi10, Utsumi13}.

\section{Instance of the convergent infinite series expression of the dynamical steady state} \la{A_S}

In Ref.\cite{Shiraishi18}, a two-level system of which Liouvillian is 
$K_{10} = 1$ and $K_{01}(\al) =\f{4\tau+1}{2\tau-\al} $ with $\al_t=t$ has been studied. 
In this case, the instantaneous steady state is given by 
$p^\RM{ss}(\al)=\big(\f{2\tau+1+\al}{4\tau+1}, \f{2\tau-\al}{4\tau+1}  \big)^t$.
The pseudo-inverse is given by
\bea
R(\al) \aeq  \begin{pmatrix} 
0 &&\f{2\tau-\al}{4\tau+1} \\
\f{2\tau-\al}{4\tau+1} &&0
\end{pmatrix} .
\eea
Then, $p^{(n+1)}(t)$ $(n=0,1,\cdots)$ becomes
\bea
p^{(n+1)}(t) \aeq  R(\al_t) \Big[\f{dR(\al_t)}{dt} \Big]^n \f{dp^\RM{ss}(\al_t) }{dt} .
\eea
The infinite series expression of the dynamical steady state \re{def_dss} converges to $\rd{p^\RM{dss}_1}(t)=\half -\f{t}{4\tau}$ and is identical with the exact solution. 

Reference \cite{Shiraishi18} set the energies as $E_0=0$ and $K_{01}(\al_t)=e^{\be E_1(t)}$. 
For $\al_t=t$ for $0 \le t \le \tau$ under the initial distribution $p_0(0)=1/2$, 
$r$ defined by \re{SFS} converges to $(5/2)\ln(3/2)=1.01366...$ for large $\tau$.

\section{Heat current and heat} \la{s_heat}

For our energy modulation \re{E(t)}, we can also interpret that the inverse temperature is time dependent $\be_t \defe \al(t)\be$. 
Then, from \re{q_1_general}, we obtain
\bea
q_{1}(t) \aeq  \be \al(t) \Big( -\big\{\al^{\pr\pr}(t)+ 2\bra E^{(0)} \ket \be[\al^\pr(t)]^2  \big\}\mu_1(\be_t) \no\\
&&+ \be[\al^\pr(t)]^2 \mu_2(\be_t) \Big), \la{q_2_general}
\eea
where 
\bea
\mu_1(\be_t)\aeqd -\sum_{i,k} E_i^{(0)} R_{ik}^{(0)} E_k^{(0)} p_k^\RM{ss} ,\\
\mu_2(\be_t)\aeqd  -\sum_{i,k} E_i^{(0)} R_{ik}^{(0)} (E_k^{(0)})^2 p_k^\RM{ss} \no\\
&&-\sum_{i,k} E_i^{(0)}  (R^{(0)}\f{dK}{d\be_t}R^{(0)})_{ik} E_k^{(0)} p_k^\RM{ss} .
\eea 
Here, $R^{(0)}_{ik}$ defined by \re{def_R} and $\sum_k R^{(0)}_{ik}(\al)p_k^\RM{ss}(\al)=0$ 
is called the Drazin inverse \cite{Drazin}.

We consider the Liouvillian \re{K_N} with fixed $\ga$ and $\beta$. 
In this case, using the Fa\`{a} di Bruno's formula \cite{Johnson}, we obtain
\bea
q_{n-1}(t) \aeq \f{(-1)^{n}}{\ga^{n-1}}\al(t) \sum \f{n!}{m_1! m_2! \cdots m_n!}  \no\\
&&\times  (-1)^{k} \be^k C_{k+1}(\be_t) \prod_j \Big(\f{\al^{(j)}}{j!} \Big)^{m_j}  , \la{q_n_R}
\eea
where $k \defe \sum m_j $ and the sum is over all $n$-tuples of nonnegative integers $(m_1, \cdots, m_n)$ satisfying the constraint $\sum_{j=1}^n j m_j  = n$. 
If $\al(t) = h_\RM{i} + \om t$ while $0<t<\tau$, \re{q_n_R} becomes
\bea
q_{n-1}(t) \aeq \f{1}{\ga^{n-1}}\al(t) \om^n \be^n C_{n+1}(\be_t) .
\eea
The heat is given by
\bea
Q_{n} \aeqd \int_0^\tau dt \ q_{n}(t) \no\\
\aeq \Big(\f{\om \be}{\ga}\Big)^{n} \f{1}{\be}\big[ \be_\RM{i} C_{n+1}(\be_\RM{i}) +C_{n}(\be_\RM{i}) \no\\
&& -\be_\RM{f} C_{n+1}(\be_\RM{f}) -C_{n}(\be_\RM{f}) \big] .
\eea

\end{document}